\documentclass{article}
\PassOptionsToPackage{square,numbers}{natbib}

\usepackage{xcolor}

\usepackage[final]{neurips_2019}

\usepackage[utf8]{inputenc} 
\usepackage[T1]{fontenc}    
\usepackage{hyperref}       
\usepackage{url}            
\usepackage{booktabs}       
\usepackage{amsfonts}       
\usepackage{nicefrac}       
\usepackage{microtype}      
\usepackage{graphicx}
\usepackage{authblk}
\usepackage{amstext}
\title{Get Real: Realism Metrics for Robust Limit Order Book Market Simulations}
\author[1]{Svitlana Vyetrenko}
\author[1,3]{David Byrd}
\author[3]{Nick Petosa}
\author[1,2]{Mahmoud Mahfouz}
\author[1]{Danial Dervovic}
\author[1]{Manuela Veloso}
\author[1]{Tucker Hybinette Balch}
\affil[1]{\footnotesize J. P. Morgan AI Research, New York, NY, USA \thanks{This paper was prepared for information purposes by the AI Research Group of J.P. Morgan Chase and Co and its affiliates, and is not a product of the Research Department of J.P. Morgan. J.P. Morgan makes no explicit or implied representation and warranty and accepts no liability, for the completeness, accuracy or reliability of information, or the legal, compliance, financial, tax or accounting effects of matters contained herein. This document is not intended as investment research or investment advice, or a recommendation, offer or solicitation for the purchase or sale of any security, financial instrument, financial product or service, or to be used in any way for evaluating the merits of participating in any transaction.} }
\affil[2]{\footnotesize Department of Electrical and Electronic Engineering, Imperial College, London, UK}
\affil[3]{\footnotesize School of Interactive Computing, College of Computing, Georgia Institute of Technology, Atlanta, GA, USA}
\begin{document}
\maketitle
\vspace{-0.5cm}
\begin{abstract}
Machine learning (especially reinforcement learning) 
methods for trading are increasingly reliant 
on simulation for agent training and testing.  
Furthermore, simulation is important for
validation of hand-coded trading strategies and 
for testing hypotheses about market structure. 
A challenge, however, concerns the robustness 
of policies validated in simulation because the
simulations lack fidelity. In fact, researchers have 
shown that many market simulation approaches fail 
to reproduce statistics and stylized facts seen in real 
markets. As a step towards addressing this we surveyed the 
literature to collect a set of reference metrics and 
applied them to real market data and simulation output.  
Our paper provides a comprehensive catalog of these
metrics including mathematical formulations where 
appropriate. Our results show that there are still significant 
discrepancies between simulated markets and real ones.
However, this work serves as a benchmark against which
we can measure future improvement.
\end{abstract}
\vspace{-0.2cm}
\section{Background and related work}
\subsection{Motivation}
Most professional investors, hedge funds, investment institutions and banks need robust means of testing trading strategies in simulation before ``going live'' with funds at risk. A key reason for this is to gain assurance that the strategy is likely to be effective. {\em Alpha} strategies aim
to profit from price movements, while {\em execution} strategies are intended to complete large volume orders while minimizing transaction costs. For instance, a pension fund may have concluded that it should reduce its holdings in a particular stock and therefore trigger a sell order for that asset. If this order were sent to an exchange as a market sell order, the price would likely fall significantly and provide the seller a lower average price than they would hope. In order to reduce transaction costs, it is a common practice to design execution strategies so that price impact is minimized by distributing a larger order as a set of smaller orders over time \cite{AlmgrenChriss}. 

Significant research effort is aimed at applying
Reinforcement Learning to a variety of trading problems
in which the learners are trained in simulation:
A reinforcement learning 
market-maker was presented in \cite{Spooner18}; a 
reinforcement learning approach to algorithmic execution 
was introduced in \cite{Nevmyvaka06}; deep hedging a 
portfolio of derivatives (including over-the-counter derivatives) in the presence of market friction was 
considered in \cite{Buehler18}; 
LSTM representations for an RL trading agent was given in \cite{Lu17}. 

In these financial trading problems, the statistical 
properties of the environment and how the market responds
to trading activity is often unknown and difficult to model. 
In such cases, repeating the process a number of times in a simulated environment allows us to eliminate the need of 
knowing transition probabilities explicitly, and an 
optimal policy can be learned from the gained 
simulated experience requiring realistic market 
simulation tools.

In real-time algorithmic trading, the actions of any 
given agent incurs response from other market participants. 
In simulation, autonomous agents can choose to place orders 
at any time and market response to them will not be 
reflected in historical data. Therefore, simple market 
replay of historical data is not sufficient for back testing 
or strategy construction. Interactive agent-based simulation (IABS) can potentially simulate interaction between 
individual market participants \cite{macal10}. In 
such simulators, price arises from incentives of 
fully autonomous agents each of whom act rationally 
in order to maximize their profits. These principles 
reflect how real markets operate; the challenge is, however, 
to find realistic agent configurations and prescribe 
agent behavior in such a way that their actions produce synthetic time series whose statistical properties 
resemble real markets. 

\subsection{Stylized Facts of Limit Order Book (LOB) Behavior}

Later sections of this paper rely on the reader's 
understanding
of the mechanisms by which electronic markets operate, so we
briefly review them here.
Public exchanges such as NASDAQ and NYSE facilitate the buying and selling of assets by accepting and satisfying buy and sell orders from multiple market participants. The exchange 
maintains an order book data structure for each asset 
traded. The LOB represents a snapshot of the supply 
and demand for the asset at a given time. It is an 
electronic record of all the outstanding buy and sell 
limit orders organized by price levels. A matching 
engine, such as first-in-first-out, is used to match 
incoming buy and sell order interest \cite{Bouchaud_book}.

\begin{figure}[!htb]
 \centering
 \includegraphics[width=0.45\textwidth]{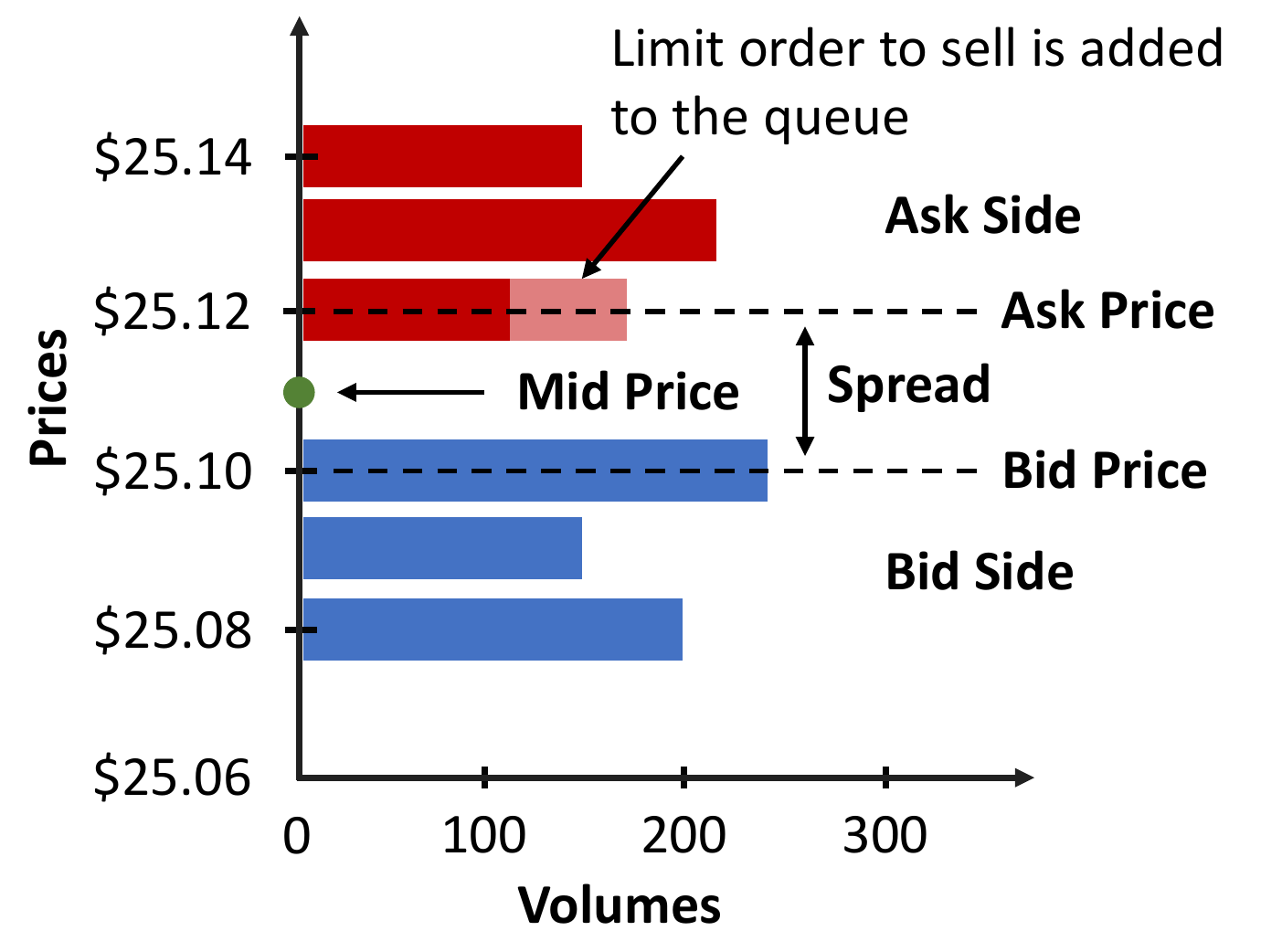}
 \caption{Visualization of the LOB structure.}
 \label{fig:LOB}
\end{figure}

Order types are further distinguished between limit orders and market orders. A limit order specifies a price that should not be exceeded in the case of a buy order, or should not be gone below in the case of a sell order. Hence, a limit order queues a resting order in the LOB at the side of the book of the market participant. A market order indicates that the trader is willing to accept the best price available immediately. A diagram  illustrating LOB structure is provided in Figure~\ref{fig:LOB}.

Many of the metrics we present below are derived from
the observation of the LOB over time.
Properties of LOB behavior that 
are repeated across a wide range of instruments, markets, 
and time periods are referred to as {\it stylized facts} \cite{cont2001empirical}. 

Evaluating the statistical properties of simulated 
asset returns, order volumes, order arrival times, 
order cancellations, etc and comparing them to those
generated from real historical data allows us to 
infer the level of fidelity of a simulation.
The question of whether stylized facts originate 
from traders' behavior,  or if they are a natural 
consequence of order book markets, has been widely 
discussed in the literature. If some stylized facts 
can be derived from markets populated only by zero 
intelligence (ZI) agents which make decisions 
without the knowledge of market microstructure, 
then these facts must originate from the mechanism 
that is governing the markets and not from strategic 
agent behavior. For example, \cite{ZI_Farmer} showed 
that a market simulation that consists only of 
ZI agents that place limit and market orders 
independently at random is able to reproduce 
price and spread dynamics as well as market 
impact. Ability of ZI agents to reproduce fat 
tails and long range dependence was shown in \cite{PalitStylizedFacts}, however, the authors
emphasized that in order to reproduce correlated 
order phenomena such as volatility clustering, 
strategic agent behavior may be required.

In this paper, we provide a survey of several 
groups of LOB stylized facts across multiple asset 
classes that lead to realism metrics with respect 
to empirical distributions (defined in Section~\ref{section_stylized_facts}). We compare 
two simulator configurations: one that contains 
zero intelligence (ZI) agents only and another 
that combines ZI agents with minimal strategic 
behavior agents. We find that the more diverse 
agent configuration behaves more similarly to 
real markets; however, we conclude that for more 
robust simulation of collective market phenomena 
online learning adaptive agents might be necessary. \cite{SchvartzmanCDAStrategies, PalitStylizedFacts}.
\vspace{-0.2cm}

\subsection{Related work}

IABS methods allow us to study phenomena that emerge 
as a consequence of multiple participant interactions 
and are difficult to model otherwise. Examples of 
such complex phenomena include both the routine 
market microstructure events such as market 
response to an individual participant's 
trading \cite{Cui_ABMImpact}; and rare events such as flash crashes \cite{LealHighFrequency, paddrik2012agent, CFTC} as well as extreme market shocks.

Wellman helped establish an empirical approach to 
the study of markets using simulated 
multi-agent systems~\cite{wellman2006methods}.
In this approach, referred to as
Empirical Game Theoretic Analysis (EGTA),
many tens of thousands of simulations are run
during which the strategies of agents are adjusted
until the system reaches a Nash equilibrium.
It is only after the system reaches equilibrium that
its statistical properties are evaluated.  
We regard this as a desirable approach, but we
have not yet implemented the methods necessary for
discovering equilibria in our experimental platform.

One notable example of multi-agent simulator use 
success is the NASDAQ tick size experiment where 
NASDAQ researchers experimentally demonstrated that 
under some agent scenarios reducing tick size would 
actually lead to increased spreads (an undesirable
property) and would negatively impact price discovery \cite{DarleyNASDAQ, BonabeauABM}. These findings 
are counter-intuitive and illustrate importance of 
multi-agent simulations for market policy research 
since one would expect that smaller tick size results 
in tighter spreads. Some explanations of these 
observations include the possibility of the existence 
and proliferation of parasitic strategies that can take advantage of better prices to make the market more 
jittery and volatile. These findings are particularly 
important since NASDAQ is interested in finding 
policies that have not yet been discovered and 
used by market participants in order to prevent 
potential market manipulations.

In real-time trading, injecting orders to the market induces other market participant activity that typically drives prices away from the agent. This activity is known as market impact \cite{AlmgrenChriss, AlmgrenPriceImpact}. Presence of market impact in real time implies that a realistic trading strategy simulation should include deviation from historical data. In literature, it is common to make an assumption of negligible market impact given the size of agent orders is small and sufficient amount of time is allowed between consecutive trades \cite{SpoonerMarketMaking}. A simple two-agent simulated market environment that consists of an algorithmic trading agent and the rest of the market is presented in \cite{VyetrenkoDecisionTrees}. It is a partially synthetic data model that allows to deviate from historical data only at times when agent places market orders which are known to cause highest market impact. This model is, however, only suited for small order placement, and is unable to capture more complex dynamics of transient price impact \cite{BouchaudTransientImpact, Gatheral2010NoDynamicArbitrageAM}. 

While modeling the market as an interplay of multiple agents seems a natural approach to mimic real market 
collective emergent behavior, justifying the realism 
of such approach for validating new trading strategies 
is difficult. Agent modeling typically relies on common sense hand-crafted rules (e.g., \cite{PalitStylizedFacts}), which can be difficult to calibrate as historical data labeled with details about each individual constituent agent behavior is typically not available for public use. Several calibration approaches---e.g. error minimization to find parameters for the asset pricing model with heterogeneous beliefs \cite{2013Tedeshi} and using Bayesian parameter estimation techniques in the simulated context---have been introduced \cite{GrazziniBayesianABM}. When individual agent- or execution strategy-specific data is available to the researcher, it can be used for the simulator calibration (e.g., \cite{VyetrenkoDecisionTrees, Yang_IRL}). 

One can view the multi-agent LOB environment as a non-cooperative game in which every agent pursues their own goal and there is no communication among agents \cite{Hu:1998}. From a game-theoretic perspective, a simulated environment is more realistic when it reaches a Nash equilibrium, where every player's parameters are configured so that each can do no better by unilaterally changing its strategy. Agents that learn to maximize their long term rewards by reinforcement from empirical equilibrium environments have been discussed in \cite{SchvartzmanCDAStrategies}.

Other approaches to IABS realism can include inverse learning agents' rewards from the market \cite{Yang_IRL}; generating synthetic LOB data using GANs \cite{li2019generating}; incorporating feedback from real-time trading into the simulation \cite{Ruiz19} and building adaptive agents that are governed by the evolutionary principles and can learn from experience \cite{LeBaron2007LongMemoryIA, Behavioral_finance}. 

\section{Realism metrics}
\label{section_stylized_facts}

We now focus on and review a series of metrics of LOB
behavior found in the literature.
One way to establish IABS realism is to ensure that 
simulated LOB time series mimic the stylized facts 
derived from real market histories. Below we review 
several groups of such stylized facts across multiple asset classes \cite{Yang_IRL, wang2017spoofing, ZI_Farmer, PalitStylizedFacts}. 

\subsection{Notation and definitions}
For simplicity of presentation, we introduce some notation and definitions that will be used throughout this paper. At time $t$, let $b_t$ be the best bid price, and let $a_t$ be the best ask price. We define mid-price as $m_t =\frac{a_t+b_t}{2}$. Choose time scale $\Delta t$. Given a time scale $\Delta t$, which can range from milliseconds to months, the log return (or simply return) at scale $\Delta t$ is defined as $r_{t, \Delta t} = \ln m_{t+\Delta t} - \ln m_t$. Let $\sigma_{\Delta t}$ be return volatility which can be calculated as standard deviation of price returns.

Let $x$ be size of a new order placed into LOB. Let $T$ be a lifetime of an order until it is fully executed or canceled. We denote by $b_t - \Delta$ the price of a new buy limit order, and $a_t + \Delta$ the price of a new sell limit order. Notice that $\Delta$ can be negative. Let $V_a$ and $V_b$ be the volumes available at the best bid and ask price. Slice LOB price and volume time series in small non-overlapping time intervals. For each time interval $\tau$, let $\mu_{V_\tau}$ be the average traded volume and $\sigma_{\tau, \Delta t}$ be the return volatility over $\tau$. Furthermore, let $P(.)$ denote probability density function of a given quantity.

\subsection{Stylized facts about asset return distributions}
Multiple stylized facts about price return distributions were studied in \cite{cont2001empirical} for equity markets as well as in \cite{BallochiEurofutures} for foreign exchange and rates markets. 
\begin{itemize}
    \item {\bf Absence of autocorrelations} Linear autocorrelations $\text{corr}(r_{t+\tau, \Delta t},r_{t, \Delta t})$ of asset returns over periods $\tau$ longer than 20 minutes are insignificant.
    \item {\bf Heavy tails and aggregational normality} The distribution of daily asset price returns shows fat tails; however, as one increases the period of time $\Delta t$ over which these returns are calculated, asset returns show lower tails. One way to quantify deviation from normal distribution is to calculate its kurtosis.
    \item {\bf Intermittency} At any micro or macro time scale, asset price returns must display high degree of volatility.
    \item {\bf Volatility clustering} High-volatility events tend to cluster in time. A quantity used to measure volatility clustering is the autocorrelation function of the squared returns $\text{corr}(r_{t+\tau, \Delta t}^2,r_{t, \Delta t}^2)$. Empirical studies using returns from various equities indicate that this autocorrelation function remains significantly positive over several days, which indicate periods of high volatility clustering \cite{cont2001empirical}.
    \item {\bf Long range dependence} If one looks at autocorrelation function of absolute returns as a function of time lag $f(\tau) = \text{corr}(|r_{t+\tau, \Delta t}|,|r_{t, \Delta t}|)$, it is empirically shown that it decays according to the power law distribution $f(\tau) \sim {\tau}^{-\beta}$ with exponent $\beta \in [0.2, 0.4]$ \cite{cont2001empirical}.
    \item {\bf Gain/loss asymmetry} Gain/loss asymmetry is prevalent for equity price returns as stocks lose value faster than they grow \cite{cont2001empirical}. However, this trend is not as pronounced for foreign exchange and rates products. Skewness is a metric that can be used to quantify the asymmetry of probability distribution about its mean.
    \item{\bf Volume/volatility positive  correlation} Volume and volatility are positively correlated. Linear regression relationship $\mu_{V_\tau} \sim \alpha + \beta \sigma_{\tau, \Delta t}$ can be derived from the data \cite{Brandouy}. 
    \item{\bf Returns/volatility negative correlation} Asset returns/volatility are negatively correlated. 
    \item{\bf Asymmetric causal information flow} Coarse-scaled volatility predicts fine-scaled volatility better than fine-scaled volatility predicts coarse scaled-volatility,
\end{itemize}

\subsection{Stylized facts about volumes and order flow}
\begin{itemize}
    \item {\bf Order book volumes} Volumes at best bid $V_b$ (and respectively volumes at best ask $V_a$) are distributed according to Gamma distribution for $\gamma \leq 1$ \cite{BouchaudStatistical}:
    $$P(V_b) \sim \exp^{-V_b} V^{-1 + \gamma}_b. $$
    \item {\bf Order sizes} Order sizes are power-law distributed \cite{Bouchaud_book}. For instance,\cite{abergel2016limit} show examples when limit order sizes are distributed as $P(x) \sim x^{-(1+\mu)}$ with exponent $1 + \mu \approx 2$ and market order sizes are distributed as $P(x) \sim x^{-(1+\mu)}$ with exponent $1 + \mu \approx 2.3-2.7$. Orders tend to have round number of shares (i.e. multiples of 10, 100, etc. are more common than neighboring sizes); in general, power-law distribution fit is product-specific. 
    \item {\bf Number of orders in a fixed time window} Number of orders in a fixed time window can be approximated by gamma or lognormal distributions \cite{abergel2016limit}.
    \item {\bf Order inter-arrival times } In the literature, LOB order inter-arrival times are suggested to be fit into exponential \cite{li2019generating}, lognormal, and Weibull distributions \cite{abergel2016limit}.
    \item {\bf New order prices } Prices at which new limit orders are placed, are power-law distributed around bid-ask \cite{abergel2016limit}. Specifically, $P(\Delta)\sim\Delta^{-(1+\mu)}$ with $1+\mu \approx 1.6$ \cite{BouchaudStatistical}.
    \item {\bf Cancellation time, time-to-first-fill and time-to-execution}
    Lifetimes of both cancelled and executed limit orders are power-law distributed, $P(T) \sim T^{-(1+\mu)}$ with $1+\mu$ ranging between $1.3$ and $1.6$ for both canceled and executed limit orders \cite{abergel2016limit}. Since an order can require multiple fills to be completed, one must distinguish between time-to-first-fill and time-to-completion statistics. Generalized gamma distribution with accelerated failure time can be used to model time-to-first-fill and time-to-completion distributions \cite{LoEconometricModels}.
    \item {\bf Time correlation of order flow} Individual agent's order placement decisions depend on other agents' actions \cite{PalitStylizedFacts}.
\end{itemize}

\subsection{Stylized facts about non-stationary patterns}
\begin{itemize}
    \item {\bf Intraday volume patterns} LOB volumes are known to exhibit strong intraday patterns. For instance, historical foreign exchange trading volumes can be approximated by fifth-degree polynomial "U-shaped" regional sessions that correspond to New York, London, and Tokyo trading \cite{Dacorogna1993AGM}. Similarly, in most equity markets, volumes are highest in the beginning of trading day, followed by a period of lower activity, and then spike again at the end of the trading day \cite{Bouchaud_book}. Note that making a transformation from physical time to tick (or transaction) time may help adjusting for intraday non-stationarity \cite{Ane2000}.
    \item {\bf Seasonal volume patterns} Some assets, especially those consumer demand for which is seasonal (e.g., electricity futures), display strong seasonal volume patterns.
    \item {\bf Intraday sensitivity to macro economic events/holidays} Due to product sensitivity to macro factors, volume spikes are known to occur in foreign exchange and rates markets during economic announcements. Equities trading is also sensitive to economic events \cite{IMF1998}. Additionally, lower trading volumes are observed on holidays throughout all asset classes.
    \item {\bf Intraday volume/spread negative correlation} Lower spreads are typically observed during periods of higher trading volumes. 
\end{itemize}

\subsection{Stylized facts about order market impact}
Market impact of order placement is a expected to grow as a function of order volume. For each time interval $\tau$, define $V_{\text{buy},\tau}$ and $V_{\text{ask},\tau}$ to be  buy and sell order volumes in $\tau$ respectively. Define participation of volume in $\tau$ as
$$
P_{\tau}=\frac{\left| V_{\text{buy},\tau} - V_{\text{ask},\tau} \right|}{V_{\text{buy},\tau} + V_{\text{ask},\tau}}.
$$
Note that $0 \leq P_{\tau} \leq 1$. Also define $\Delta m_\tau$ to be the observable mid-price move in $\tau$. Discretize the range for $P_{\tau}$ into bins $B_i, i=1, \ldots, N$ such that $B_i =\{\tau: \frac{i-1}{N} \leq P_{\tau}\leq \frac{i}{N} \}$. For each $B_i$, define 
\begin{eqnarray*}
M_i = \frac{1}{|B_i|}\sum_{\tau \in B_i} \Delta m_\tau 
\quad\text{and}\quad 
P_i = \frac{1}{|B_i|}\sum_{\tau \in B_i} \Delta P_\tau 
\end{eqnarray*}
to be the average price move and average participation of volume in bins with similar volume participation. One can then fit the relationship of the form $M_i \sim \alpha P_i ^{\beta}$ through the data \cite{AlmgrenPriceImpact, ZI_Farmer, bouchaud2010price}.

\subsection{Stylized facts about cross asset correlations}
When simulating multiple assets, cross asset correlation properties must hold. For instance, equity index and its major constituents must show high degree of correlation \cite{cont2001empirical}. For futures, for example, asset price moves across term structure are highly correlated and exhibits consistent patterns uncovered by the PCA (e.g., \cite{BallochiEurofutures}). It is also worth noting that extreme returns (e.g., 99-percentile returns that occur during financial crisis) across various stocks or asset classes can be extremely correlated while their average returns are not  \cite{cont2001empirical}.

\section{Experiments with a multi-agent simulation}

\subsection{ABIDES - Agent-Based Interactive Discrete Event Simulation environment}
In order to evaluate the ability of a given agent configuration to reproduce stylized facts about the market, we employ ABIDES, an agent-based interactive discrete event simulation environment \cite{simulator}. ABIDES provides a selection of background agent types (such as agent types described in Section~\ref{sec_agent_types}), a NASDAQ-like exchange agent which lists any number of securities for trade against a LOB with price-then-FIFO matching rules, and a simulation kernel which manages the flow of time and handles all inter-agent communication.  Trading agents may not inspect the state of the exchange directly, but must direct realistic messages to request order book depth, obtain last trade prices, or place or cancel limit orders through the kernel, which imposes delays for computational effort and communication latency.  Time proceeds in nanoseconds and all securities are priced in cents. The ABIDES source code is freely available under a BSD-style license at \texttt{https://github.com/abides-sim/abides}.

\subsection{Background agent types}
\label{sec_agent_types}
In order to conduct experiments, we specify one ZI and three minimal strategic agent types (market maker, momentum and heuristic belief learning agents).  Detailed explanations and examples for the ZI and HBL agents and the fundamental process which drives them are available in \cite{byrd2019explaining}.

{\bf Zero intelligence agents:} 
Originally introduced in  \cite{gode1993allocative}, ZI agent class includes a variety of agents that do not base their trading decisions on the knowledge of LOB microstructure. Similarly to \cite{wah2017welfare}, ZI agents in our implementation are enhanced with knowledge of noisy observation of exogenous 'true' value of stock which represents agent's understanding of the outside world (eg. earnings reports, macro events, immediate trading demand). ZI agents in our experiment arrive to the market according to a Poisson process and do not use LOB microstructure signals for trading decisions. 

\textbf{Heuristic belief learning (HBL) agent:}
HBL agents base their decisions on a limited-length historical snapshot of the order stream, which they use to maximize expected surplus using a heuristic estimation of the probability that a given limit price will successfully transact in the market \cite{gjerstad2007competitive}, \cite{gjerstad1998price}. Our implementation matches that of \cite{wang2017spoofing}.

{\bf Market maker agent:} The market maker agent acts as a liquidity provider by placing orders on both sides of LOB with a constant arrival rate of 10 seconds. The agent starts by cancelling any existing orders. It then queries the current spread to determine the prices of the buy and sell orders to be submitted. The agent is configured to place orders on top $N$ levels on both sides of LOB with the size split determined by the number of price levels it quotes and the order volume chosen uniformly at random within fixed bounds. Our implementation is similar to that of \cite{McGroarty2018HighFT, chakraborty2011market}.

{\bf Momentum agents:} The momentum agents base their trading decision on observed price trends. Our implementation compares the 20 past mid-price observations with the 50 past observations and places a buy order of random size, if the former exceeds the latter and a sell order otherwise. 
\vspace{-0.2cm}

\subsection{Agent configurations}
\label{agent_configuration}
In this paper we present two background agent configurations, one that is composed exclusively of ZI agents (\textit{sparse\_zi\_100}---contains 100 ZI agents), and one that has both ZI and strategic agents (\textit{rmsc01}---contains 1 market maker agent, 50 ZI agents, 25 HBL agents and 24 momentum trading agents). Note that in both configurations, all agents are permitted to reenter the market. To assess the realism of \textit{sparse\_zi\_100} and \textit{rmsc01} ABIDES configurations, we look at stylized facts about asset returns and order flow described in Section~\ref{section_stylized_facts}.

\section{Experimental results}
\subsection{Stylized facts about asset return distributions}
To derive historical asset return distributions, we analyze minutely intraday log returns of 50 randomly sampled U.S. exchange-traded equities for each trading day of 2011. The set of equities is resampled each trading day and is drawn uniformly across all stocks from all exchanges.

\begin{itemize}

\item{\bf Heavy tails and aggregation normality.}
We confirm experimentally that both \textit{rmsc01} and \textit{sparse\_zi\_100} show heavy tails and aggregational normality, with \textit{rmsc01} being closer to historical data (see Figure~\ref{fig:ReturnsDist}). 

\begin{figure}[!htb]
   \centering
   \includegraphics[width=0.35\textwidth]{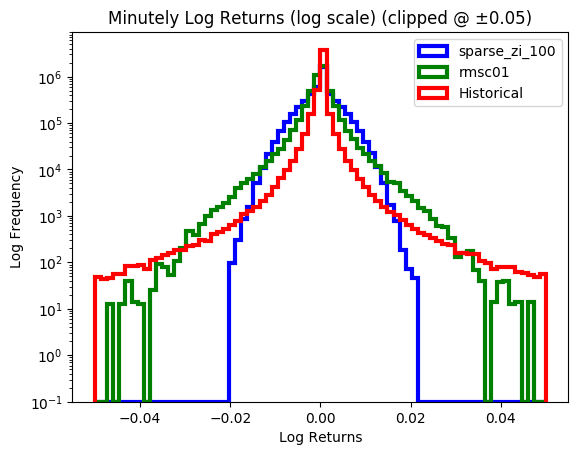}
   \includegraphics[width=0.35\textwidth]{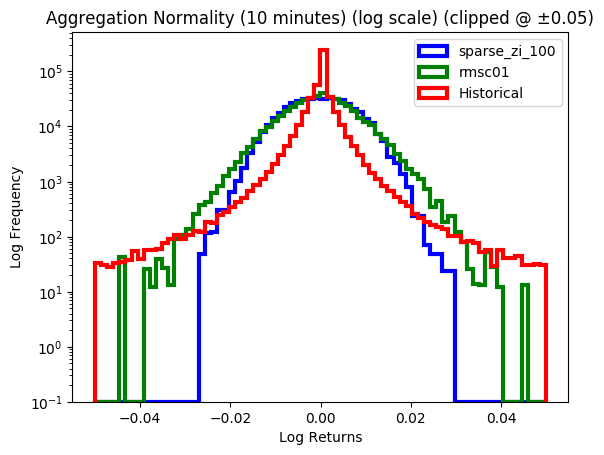}
   \caption{One-minute (left) and ten-minute (right) log return distributions.}
   \label{fig:ReturnsDist}
   \end{figure}

\item{\bf Absence of autocorrelations}
Figure~\ref{fig:three} ({\it left}) shows the correlation coefficient distributions for lag 1 autocorrelation of log returns over 30 minute intervals. Historical market data exhibits a notable spike of 0 correlation, in line with previous discussion in Section~\ref{section_stylized_facts}. Both the \textit{sparse\_zi\_100} and \textit{rmsc01} agent configurations fail to capture this stylized fact. When comparing the overall distribution shape, the \textit{rmsc01} configuration resembles the shape of historical autocorrelation distribution closer.

\item{\bf Volatility clustering.}
The average autocorrelation of square returns decays for both historical and simulated data as time lag increases (see Figure~\ref{fig:three} ({\it center})). 
\item{\bf Volume/volatility correlation.}
Neither simulated distribution of correlation coefficients between volume and volatility was as skewed or high-variance as that of historical data (see Figure~\ref{fig:three} ({\it right})).

\begin{figure}[!htb]
   \centering 
   \includegraphics[width=0.32\textwidth]{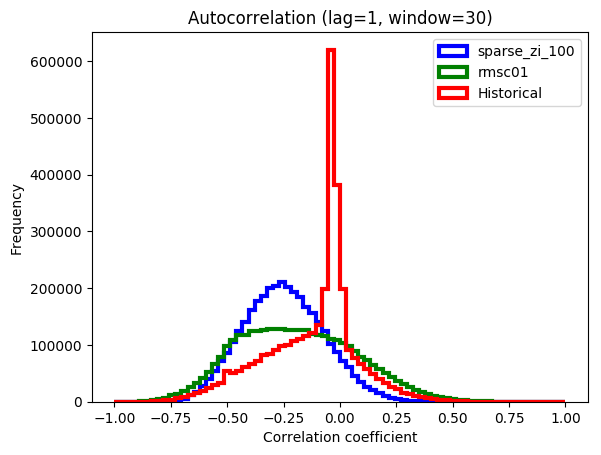}
   \includegraphics[width=0.32\textwidth]{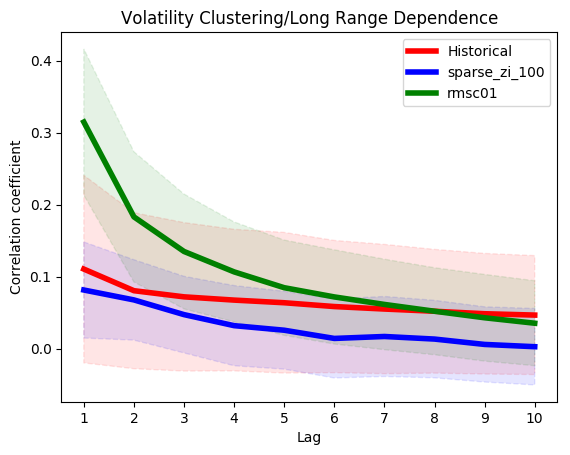}
   \includegraphics[width=0.32\textwidth]{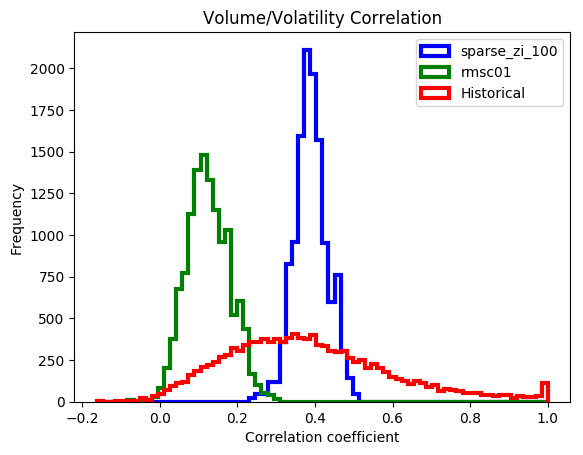}
   \caption{
   ({\it left}) Distributions of return autocorrelation. ({\it center}) Average autocorrelation of square returns as a function of time lag. ({\it right}) Volume/volatility correlation distributions.
   }
   \label{fig:three}
   \end{figure}


\end{itemize}

\subsection{Stylized facts about volumes and order flow}

To derive historical distributions, we consider order book historical data for JPM stock traded on the NASDAQ exchange for each trading day of June 2019 from 9:30 am to 4:30 pm.

\begin{figure}[!htb]
   \centering 
   \includegraphics[width=0.32\textwidth]{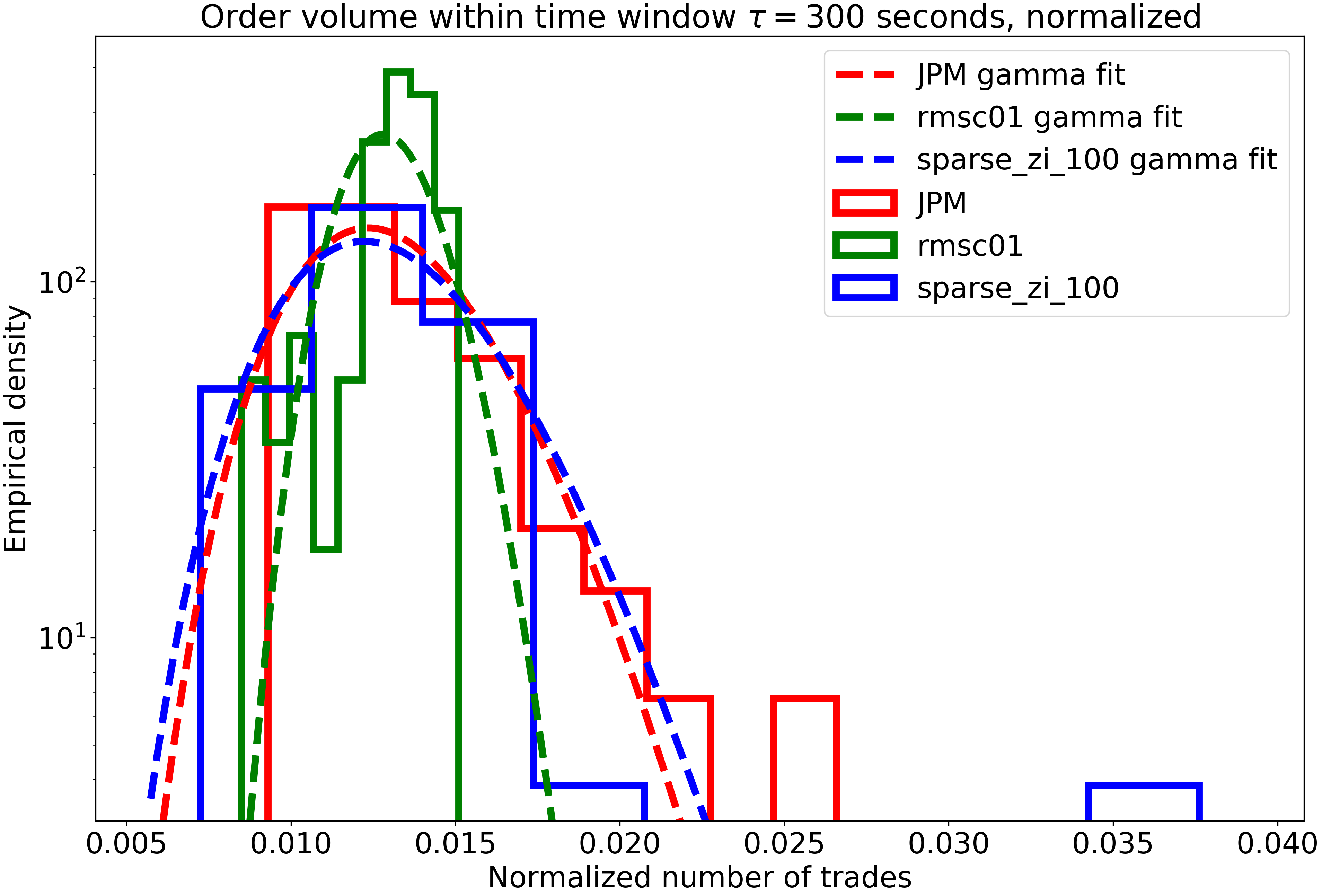}
   \includegraphics[width=0.32\textwidth]{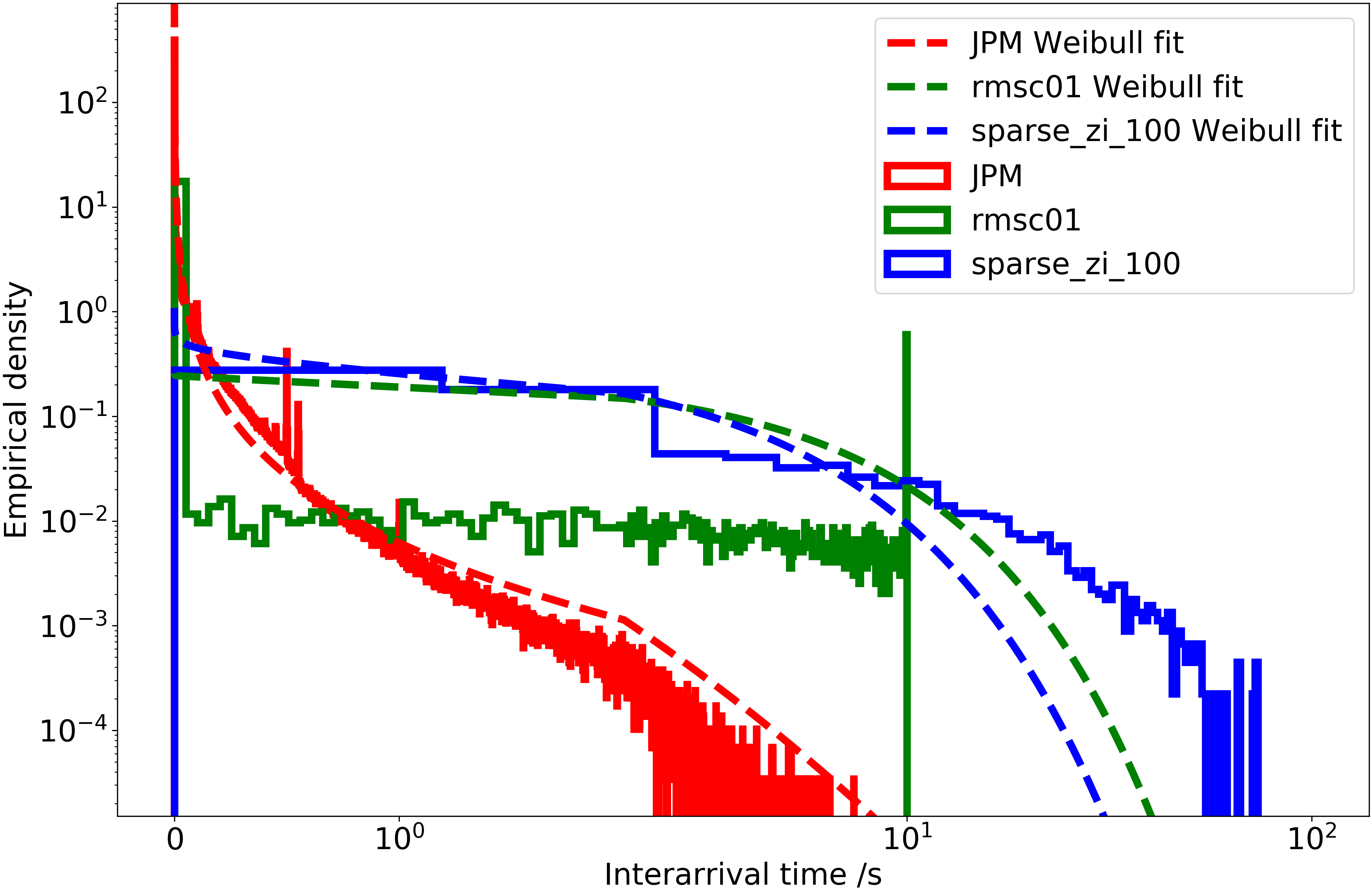}
   \includegraphics[width=0.32\textwidth]{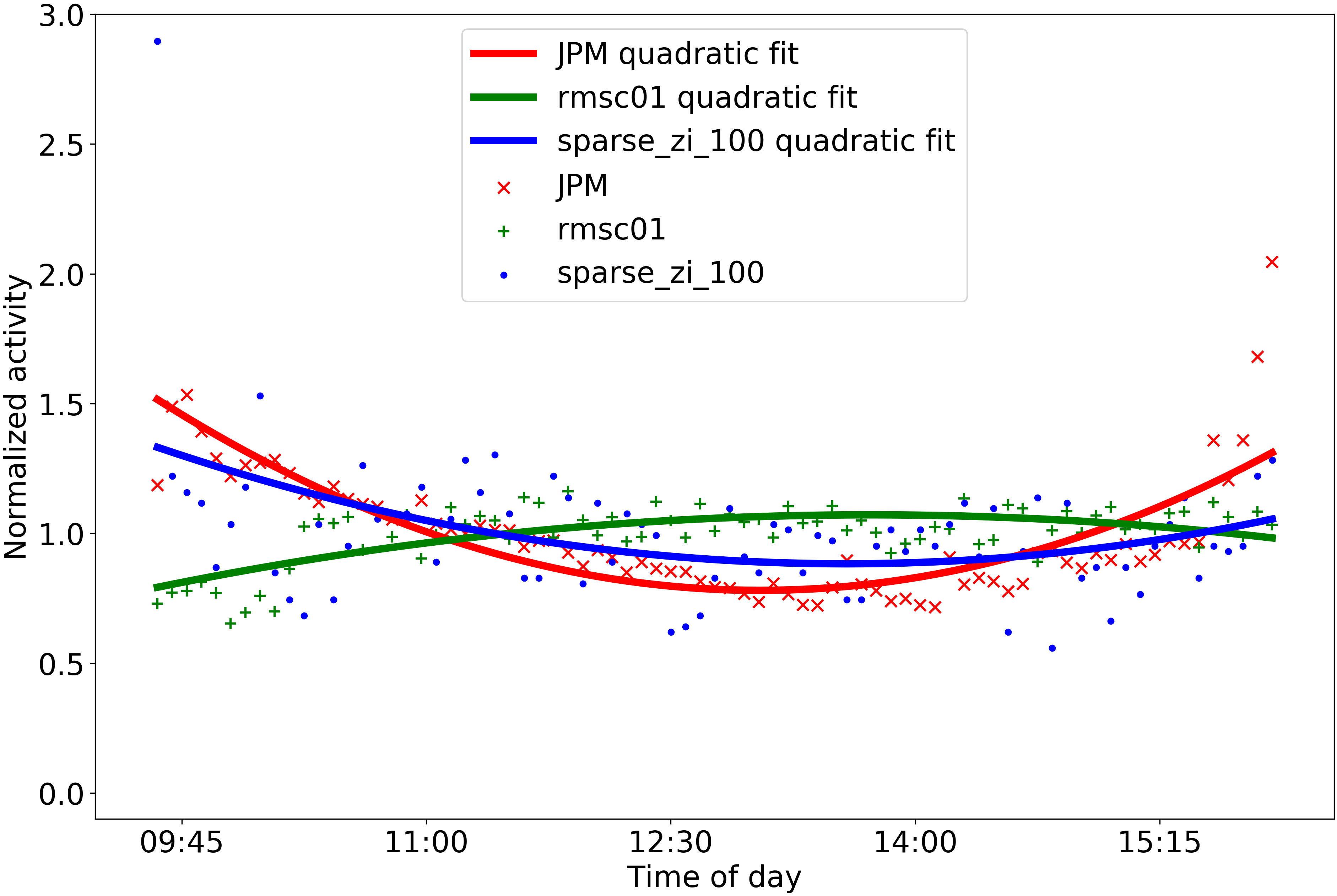}
   \caption{
   ({\it left}) Number of orders in a five-minute window. ({\it center}) Limit order interarrival times. ({\it right}) Intraday volume profiles.
   }
   \label{fig:order_flow}
   \end{figure}

\begin{itemize}

    \item{\bf Number of orders in a fixed time window.}
    Figure~\ref{fig:order_flow} ({\it left}) shows limit order volume distribution in a five-minute window for the simulated vs. the historical data. We find that gamma distribution produces a good fit for these curves.
    
    \item{\bf Intraday volume patterns.} 
    Quadratic curves have been fitted to this data to demonstrate the "U-shaped" pattern of historical intraday volumes (see Figure~\ref{fig:order_flow} ({\it right}) ). The simulation data for \emph{rmsc01} shows the reverse of expected activity: namely, reduced trading at the open and close of the market.

    \item{\bf Order interarrival times.}
    Figure~\ref{fig:order_flow} ({\it right}) shows distribution of interarrival times for simulated data as well as the historical data. Historical data fits Weibull distribution rather well. The distribution for \emph{rmsc01} is dominated by values close to zero resulting in a poor fit to the Weibull distribution.
    \end{itemize}
\vspace{-0.2cm} 
\vspace{-0.3cm}
\section{Conclusion and Discussion}
In this paper, we provided a catalog of known stylized facts regarding
LOB microstructure behavior with respect to market realism. 
We chose ABIDES environment \cite{simulator} as a simulation platform and evaluated two experimental configurations of
agent types---one exclusively of ZI agents and 
one with ZI agents and other agents with minimal 
strategic behavior. 

We observed that the configuration with
a more diverse agent population leads to statistics
that more closely mimic real markets, we acknowledge 
that there is much room for improvement. 
In particular, time correlation of order flow and intraday volume patterns are not well reproduced. While additional 
constraints might be needed to produce stylized 
facts about intraday patterns from agents' incentives 
(e.g., require agents to close all positions by the end of the trading day), correlated order behaviors, especially 
herding or clustering behaviors, require adaptation 
of one agent's behavior in response to other agents' 
actions and will possibly require introduction of 
online learning agents \cite{PalitStylizedFacts}.  
For example, \cite{LeBaron2007LongMemoryIA} conducted comparisons of non learning and learning agents and 
concluded that agents capable of learning and 
adaption to other agent flows are able to replicate 
stylized facts about long range dependence and correlation between volume and volatility better. Specifically, during periods of high volumes, when more agents are trading in response to others' behavior, higher trading activity keeps volume queues available at best bid or ask levels relatively short; hence, LOB layers  move more frequently and, as a result, prices are more volatile. Moreover, in real markets, rational agents evolve over time by learning to expand effective and cull ineffective trading strategies \cite{Behavioral_finance}. Hence, we believe that enhancing autonomous LOB agents with ability to learn from experience will be a step towards making simulated environments more robust.
\vspace{-0.2cm}
\section{Acknowledgements}
This material is based on research supported in part by the National Science Foundation under Grant no. 1741026, and by a J.P.Morgan AI Research Fellowship.
\newpage
\bibliography{bib}
\bibliographystyle{unsrt}

\end{document}